\documentclass[12pt]{article}
\usepackage[margin=1in]{geometry}
\usepackage{amsmath}
\usepackage{times}
\usepackage{graphicx}
\usepackage{color}
\usepackage{rotating}
\usepackage{bbm}
\usepackage{latexsym}

\usepackage{epsfig}
\usepackage{amssymb}
\usepackage{algorithm}
\usepackage{algorithmic}
\usepackage{slashbox}

\begin{document}
\begin{center} \textbf{ \Large Bayesian estimators of the Gamma distribution} \end{center} 
\begin{center} A. Llera, C. F. Beckmann.\end{center} 
\begin{center} {\footnotesize Technical report \\ Radboud University Nijmegen \\ Donders Institute for Brain Cognition and Behaviour.} \end{center} 
\vspace{0.5 cm}
\begin{center} {\bf Abstract} \end{center}
In this paper we introduce two Bayesian estimators for learning the parameters of the Gamma distribution. The first algorithm uses a well known unnormalized conjugate prior for the Gamma shape and the second one uses a non-linear approximation to the likelihood and a prior on the shape that is conjugate to the approximated likelihood. In both cases use the Laplace approximation to compute the required expectations. We perform a theoretical comparison between maximum likelihood and the presented Bayesian algorithms that allow us to provide non-informative parameter values for the priors hyper parameters. 
We also provide a numerical comparison using synthetic data. The introduction of these novel Bayesian estimators open the possibility of including Gamma distributions into more complex Bayesian structures, e.g. variational Bayesian mixture models.

\vspace{0.5 cm}
\section{Introduction}
The Gamma distribution is the most considered when modeling positive data \cite{Lit:Gm_2,Lit:Gm_1,minka2002estimating,Lit:Gm-N-Gm_1}. The fastest and oldest method used to estimate the parameters of a Gamma distribution is the Method of Moments (MM) \cite{Lit:Gm_2}. This method can be applied to the Gamma distribution since there exists an unique relationship between its first two moments and its parameters. The MM uses a Gaussian approximation to the moments to provide a closed form parameters estimation. Obviously this method is exact for the Gaussian distribution and less accurate for distributions deviating from Gaussianity. Another classic approach for parameter estimation is the well known maximum likelihood (ML), based in the maximization of the data log-likelihood; for the Gamma distribution there exist a closed form ML derivation for its scale parameter but not for the shape parameter. Two interesting algorithms to deal with this issue can be found in \cite{minka2002estimating}. One of them uses a linear approximation to the otherwise intractable terms appearing in the derivative of the log-likelihood and the second one uses a more advanced non-linear approximation to the log-likelihood to obtain higher convergence rate \cite{Minka_newton,minka2002estimating}. While the MM and ML approaches provide point estimates for the distribution parameter values, a full Bayesian estimation approach introduces distributions over the parameters \cite{MB}; when choosing conjugate prior distributions over the parameters \cite{MB,bishop}, the posterior expectations can be analytically derived. The conjugate prior for the Gamma rate parameter is known to be Gamma distributed but there exist no proper conjugate prior for the shape parameter. 
  
In section \ref{bayes} we present the Bayesian methods proposed in this article. In subsection \ref{b:sc} we review the conjugate prior for the rate (inverse scale) parameter. Since for the shape parameter there exist no proper conjugate prior distribution, here we propose two different strategies to deal with this issue. In subsection \ref{b:sh} we consider the use of an unnormalized conjugate prior \cite{Fink97acompendium} and in section \ref{IG_shape_post2} we introduce an unnormalized prior on the shape which is conjugate with the accurate non-linear likelihood approximation presented in \cite{Minka_newton}. In both cases we use the Laplace approximation to compute the required expectations. In section \ref{res:st} we provide a theoretical comparison between the two ML algorithms presented in \cite{minka2002estimating} and the two presented Bayesian strategies. In section \ref{num_res} we present a numerical comparison between the four algorithms. In section \ref{res:sh} we briefly consider the hyper parameters value choices for the Bayesian algorithms.  We conclude the paper with a brief discussion in section \ref{dis}. For completeness of the methodology and ease of reading, in Appendix \ref{MM} we review the Method of Moments estimation, and in Appendix \ref{ML} the two Maximum Likelihood algorithms introduced in \cite{minka2002estimating}.

\section{Bayesian Learning (BL) for the Gamma distribution}
\label{bayes}
Consider the Gamma distribution defined over the support $x \in \mathbb{R}^+$ 
\begin{align}
\label{ig}
\mathcal{G}(x|\alpha,\beta)= \frac{ x^{\alpha -1}}{\Gamma(\alpha) \beta^{\alpha}} \exp(\frac{-x}{\beta}),
\end{align} 
for any positive real parameter values $\alpha$ and $\beta$, denoting the distribution shape and scale respectively, and where $\Gamma$ is the Gamma function. Given a vector of positive real values $\textbf{x}=\{x_1,\ldots,x_n \}$, in the following subsections we present two different Bayesian strategies to find posterior estimations $\hat{\alpha}$ and $\hat{\beta}$ for the distribution parameters $\alpha$ and $\beta$. 

To find the posterior probability of the Gamma parameters, $\theta=\{\alpha, \beta \}$, we use Bayes rule 
\begin{align}
\label{by}
p(\theta|\textbf{x}) = \frac{ p(\textbf{x}|\theta) p(\theta)}{p(\textbf{x})},
\end{align} 
where $\textbf{x}= \{x_1,\ldots,x_n \}$ is some positive vector of observations. Since the denominator only depends on data the posterior is proportional to the likelihood multiplied by the prior
\begin{align}
\label{by}
p(\theta|\textbf{x}) \propto  p(\textbf{x}|\theta) p(\theta).
\end{align} 
Obtaining analytical solutions for the parameters $\theta$ requires the use of conjugate priors \cite{bishop,Fink97acompendium}. A prior is called conjugate with a likelihood function if the prior functional form remains unchanged after multiplication by the likelihood. 

\subsection{Rate parameter $R=\frac{1}{\beta}$ }
\label{b:sc}
A well known conjugate prior for $R=\frac{1}{\beta}$, the rate parameter of the Gamma distribution, is a Gamma distribution parametrized using shape $d$ and rate $e$ \cite{Fink97acompendium}, 
\begin{align}
\label{pb}
p (R)=  \mathcal{G}(R|d,e). 
\end{align}

Given the observations vector $\textbf{x}$, and multiplying its Gamma likelihood by the prior on the rate (\ref{pb}) we get its posterior, $q(R)= \mathcal{G}(R|\hat{d},\hat{e})$ with
\begin{align}
\label{td}
\hat{d}= d + n \alpha,    \hspace{2cm} \hat{e}=e+\sum_{i=1}^{n} x_i.
\end{align} 
Since $R$ is Gamma distributed its posterior expectation is
\begin{align}
\label{beta}
\hat{R} =  \frac{\hat{d}}{\hat{e}}.
\end{align} 
Since the scale is the inverse of rate we have that $\hat{\beta}=\frac{\hat{e}}{\hat{d}}$. 

\subsection{Unnormalized prior on shape parameter $\alpha$ (BL1)}
\label{b:sh}
The unnormalized prior used for the shape parameter of the gamma distribution \cite{Fink97acompendium} has the form
\begin{align}
\label{IG_shape_prior}
p (\alpha) \propto \frac  {a^{\alpha-1} R^{\alpha c}}   {\Gamma(\alpha)^{b}},
\end{align}
where $R$ is the Gamma rate parameter and $\{a,b,c\} \in \mathcal{R}^{+}$ are hyper parameters. 
Given some observations $\textbf{x}$, we multiply its likelihood under the Gamma distribution by the prior on shape (\ref{IG_shape_prior}) to obtain an expression for the posterior distribution $q(\alpha)$
\begin{align}
\label{IG_shape_post}
q (\alpha) \propto \frac  {\hat{a}^{\alpha-1} R^{\alpha \hat{c}}}   {\Gamma(\alpha)^{\hat{b}}}
\end{align}
with 
\begin{align}
\label{ha}
\hat{a}= a \prod_{i=1}^{n} x_i, \hspace{2cm} \hat{b}=b+n, \hspace{2cm} \hat{c}= c + n. 
\end{align} 
Finding the posterior expectation of $\alpha$, $\hat{\alpha}$, implies computing the expectation of $p(\alpha)$. Here we use the Laplace approximation to (\ref{IG_shape_prior}), which can be shown to be a Gaussian with mean 
\begin{align*}
m= \Psi^{-1}\Big(\frac{\log a + c \log R}{b}\Big),
\end{align*}
and precision $b\Psi_1(m)$, where $\Psi_{1}(\alpha) =\frac{d \Psi(\alpha)}{d\alpha}$.
Consequently we approximate the posterior expected shape by
\begin{align}
\label{mu22}
\hat{\alpha} \approx \Psi^{-1}\Big(\frac{\log \hat{a} + \hat{c} \log R}{\hat{b}}\Big).
\end{align}
We note here that the expectation of the Laplace approximation to $q(\alpha)$ corresponds to the maximum a posteriori (MAP) estimate of $q(\alpha)$. The use of the Laplace approximation in this context then reduces to using a MAP estimation in place of the expected value. 
Note also that in order to compute the expectation (\ref{mu22}) we need to estimate $\log \hat{a}$ 
\begin{align}
\label{lha}
\log \hat{a}= \log a + \sum_{i=1}^{n} \log x_i,
\end{align} 
and not longer require $\hat{a}$ in equation (\ref{ha}). This fact has the advantage of avoiding numerical issues for large sample sizes. Further, substituting $\hat{R}= \frac{\hat{d}}{\hat{e}}=\frac{d + n\alpha}{e+\sum_{i=1}^{n} x_i}$ into equation (\ref{mu}) we obtain an expression with no $R$ dependence which is more compact for algorithmic use, namely
\begin{align}
\label{mu2}
\hat{\alpha} \approx \Psi^{-1}\Big(\frac{\log \hat{a} + \hat{c} \big( \log (d + n\alpha) -\log (e+\sum_{i=1}^{n} x_i)   \big)}{\hat{b}}\Big).
\end{align}

\begin{algorithm}
\label{aa100}
\caption{BL1}
\begin{algorithmic}
\REQUIRE $\textbf{x}=\{x_1,\ldots,x_n \}, x_i > 0$ and $\{a, b, c, d, e\}$ 
\STATE  $\mu= \frac{1}{n} \sum_{i=1}^{n} x_i$
\STATE  $v= \frac{1}{n-1} \sum_{i=1}^{n} (x_i - \mu)^2$
\STATE $\alpha = \frac{\mu^2}{v}$
\STATE $\hat{e}=e+\sum_{i=1}^{n} x_i$
\STATE  $\log \hat{a}= \log a + \sum_{i=1}^{n} \log x_i$
\STATE $\hat{b}=b+n$
\STATE $\hat{c}= c + n$
\REPEAT
\STATE $\alpha \leftarrow \Psi^{-1}\Big(\frac{\log \hat{a} + \hat{c} \big( \log (d + n \alpha)    -\log (\hat{e})  \big)}{\hat{b}}\Big)$
\UNTIL {convergence}
\STATE $\hat{\alpha}=\alpha$
\STATE $\hat{d}= d + n \hat{\alpha} $
\STATE $\hat{\beta}=\frac{\hat{d}}{\hat{e}}$
\RETURN $\hat{\alpha}$,$\hat{\beta}$
\end{algorithmic}
\end{algorithm}
Algorithm 1 summarizes this first Bayesian approach for learning the Gamma parameters which we denote as $BL1$. First, the MM algorithm is used to initialize $\alpha$ (equation (\ref{mmig}) left panel); $\hat{e}$, $\log \hat{a}$, $\hat{b}$ and $\hat{c}$ are computed through equations (\ref{td} right side), (\ref{lha}) and (\ref{ha}). Then equation (\ref{mu2}) is iterated till convergence to obtain the expected $\hat{\alpha}$. Finally $\hat{d}$ is computed using equation (\ref{td} left side) and 
$\hat{\beta}$ through  (\ref{beta}).

\subsection{Likelihood approximation and its conjugate prior on $\alpha$ (BL2)}
\label{IG_shape_post2}
Given an initial parameter value estimation $\alpha$, we use an approximation to the log-likelihood as presented in \cite{Minka_newton}, namely 
\begin{align}
\label{log_likkk}
\log \mathcal{G}(\mathbf{x}|\alpha)  = k_0 + k_1 \alpha + k_2 \log \alpha.
\end{align} 
Learning the parameters $k_0,k_1,k_2$ is straightforward (check Appendix B). We then define a prior on $\alpha$ that is conjugate with this approximated log-likelihood, namely
\begin{align}
\label{log_cp}
\log p(\alpha) \propto w_0 + w_1 \alpha + w_2 \log \alpha
\end{align} 
for any set of hyper parameter values $w_0,w_1,w_2$. It is straightforward to observe that the posterior values for the hyper parameters are 
\begin{align}
\tilde{w}_i=w_i +k_i, \hspace{1cm}  \forall i \in \{0,1,2 \}.
\end{align} 
Once more we use the Laplace approximation to (\ref{log_cp}), which can be shown to be a Gaussian with mean 
\begin{align*}
m= \frac{-\tilde{w_2}}{\tilde{w_1}}
\end{align*}
and precision $\frac{\tilde{w_2}}{\alpha^2}$. 
Consequently we approximate the posterior expected shape by
\begin{align}
\label{mu}
\hat{\alpha} \approx \frac{-\tilde{w_2}}{\tilde{w_1}}.
\end{align}
Again, the Laplace approximation corresponds to the maximum a posteriori (MAP) estimate. Algorithm 2 summarizes this alternative for Bayesian learning the parameters of the Gamma distribution which will be further denoted as $BL2$.

\begin{algorithm}
\label{aa1000}
\caption{BL2}
\begin{algorithmic}
\REQUIRE $\textbf{x}=\{x_1,\ldots,x_n \}, x_i > 0$ and $\{w_1,w_2\}$ 
\STATE  $\mu= \frac{1}{n} \sum_{i=1}^{n} x_i$
\STATE  $v= \frac{1}{n-1} \sum_{i=1}^{n} (x_i - \mu)^2$
\STATE $\alpha = \frac{\mu^2}{v}$

\REPEAT
\STATE $k_1= n \big( \overline{\log \textbf{x}} - \Psi(\alpha) - \log \overline{\textbf{x}} + \log \alpha - \alpha \Psi'(\alpha) +1\big)$
\STATE $k_2= n \alpha^2 \Psi'(\alpha) - n \alpha$ 
\STATE $\tilde{w_1}= w_1+k_1$
\STATE  $\tilde{w_2}= w_2+k_2$
\STATE $\alpha \leftarrow \frac{-\tilde{w_2}}{\tilde{w1}}$
\UNTIL {convergence}
\STATE $\hat{\alpha}=\alpha$
\STATE $\hat{d}= d + n \hat{\alpha} $
\STATE $\hat{\beta}=\frac{\hat{d}}{\hat{e}}$
\RETURN $\hat{\alpha}$,$\hat{\beta}$
\end{algorithmic}
\end{algorithm}

\section{Results}
\label{results}
In section \ref{res:st} we highlight the relationship between the ML algorithms presented in \cite{Minka_newton} and the Bayesian ones presented here. 
Both ML algorithms (ML1 and ML2) can be found in Appendix \ref{ML}. In section \ref{num_res} we present numerical results obtained using synthetic data and 
in section \ref{res:sh} we consider the hyper parameter value setting for the Bayesian shape priors. 
\subsection{Algorithms comparison}
\label{res:st}
It is to note that ML and BL algorithms have notable similarities; first, note that the ML and the Bayesian scale estimators presented are 
\begin{align}
\hat{\beta}_{ML}= \frac{\mu}{\alpha}=\frac{ \sum_{i=1}^{n} x_i}{n \alpha} \hspace{2cm} \hat{\beta}_{BL}= \hat{R}^{-1}= \frac{\hat{e}}{\hat{d}}=\frac{e+\sum_{i=1}^{n} x_i}{d + n\alpha}
\end{align}
respectively. It is easy to observe that 
\begin{align}
\lim_{d=e \to 0} \hat{\beta}_{BL} \rightarrow \hat{\beta}_{ML},
\end{align}
which means that the Bayesian scale estimator tends to the ML one in the limit of an infinite variance gamma prior over $\beta$. With respect to the shape parameter $\alpha$, 
since initializing $b=c$ results in $\hat{b}=\hat{c}$ (in BL1), and taking again the same limit, we can rewrite the $BL1$ $\alpha$ update as
\begin{align}
\lim_{ \substack{d=e \to 0 \\ b=c}} \hat{\alpha}_{BL} \rightarrow \Psi^{-1}\Big(\frac{\log \hat{a}}{\hat{b}} + \log (n \alpha) -\log \sum_{i=1}^{n}x_i  \Big)
\end{align}
Comparing this to the ML $\alpha$ update in the case of a linear constrain on $\alpha$ (ML1)
\begin{align}
\hat{\alpha}_{ML1} = \Psi^{-1}(  \log \alpha + \overline{\log \textbf{x}} - \log \bar{x} )=  \Psi^{-1}( \overline{\log \textbf{x}} + \log n\alpha - \log \sum_{i=1}^{n}x_i )
\end{align}
it is clear that both models perform an iterative update of the shape parameter dependent in the inverse of the gamma function of $\log n \alpha$ plus a data dependent constant which differs for ML1 and BL1; they also have the same dependence on one of the data dependent terms, $-\log \sum_{i=1}^{n} x_{i}$, independently of the hyper parameter values used for the prior on $\alpha$ (under the assumption b=c). Further, since $\log \hat{a}= \log a + \sum_{i=1}^{n} \log x_i$ and $\hat{b}=b+n$, in the extreme case of considering a very small hyper parameter value for $b$, we can simplify the BL1 estimation update further, resulting in 
\begin{align}
\lim_{ \substack{d=e \to 0 \\ b=c \to 0}} \hat{\alpha}_{BL} \rightarrow \Psi^{-1}\Big(\frac{\log a}{n}  + \overline{\log \textbf{x}} + \log (n \alpha) -\log \sum_{i=1}^{n}x_i  \Big)
\end{align}
Remembering that the posterior estimation for $b$ and $c$ are $\hat{b}=b+n$ and $\hat{c}=c+n$, this is a very interesting result that shows that when choosing little informative hyper parameter values for $b$ and $c$, the BL1 estimation involves a small sample bias correction term $\frac{-\log a}{n}$ which tends to $0$ as the number of observed samples increases.  
Further, there is, by construction, a close relationship between ML2 and BL2; in fact they are equivalent when considering a flat prior over $\alpha$, that is when $w_1=w_2=0$ in equation (\ref{log_cp}).

\subsection{Numerical results}
\label{num_res}
\begin{figure}[!]
\centering  	
\includegraphics[width=.9\textwidth]{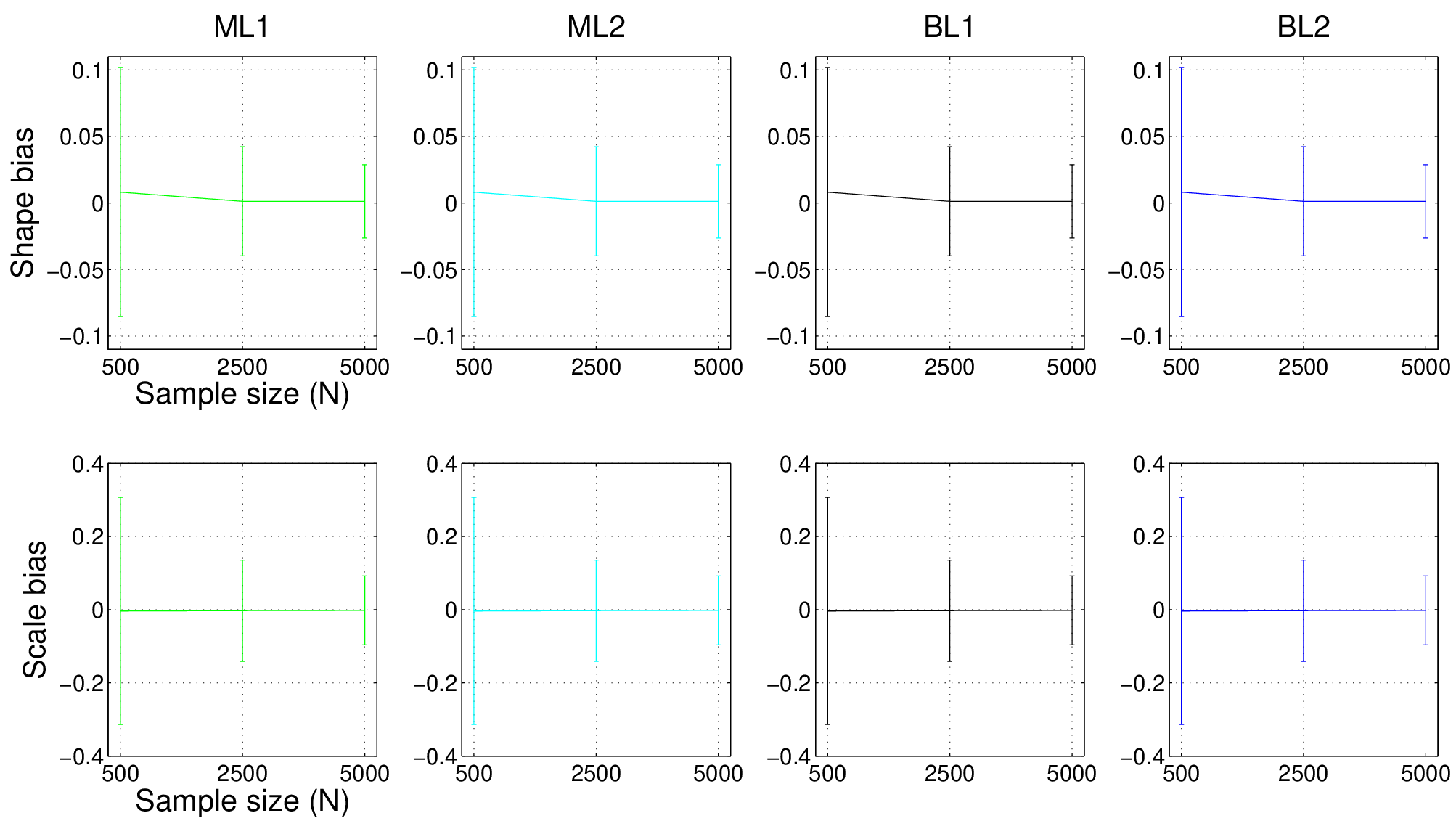}.
\caption{.} 
\label{fig:bias}
\end{figure}

In this section we perform a numerical comparison between the two maximum likelihood (ML) algorithms presented in \cite{minka2002estimating} and the Bayesian approaches (BL) presented in this work. For each simulation we generate varying amount of $N$ samples from a Gamma distribution with fixed parameters $\alpha$ and $\beta$ and we applied the four considered algorithms. The parameter values $\alpha$ and $\beta$ are initialized to different random positive numbers at each simulation. For each $N$ we performed 500 different simulations. The ML and the BL algorithms are considered to converge when the relative $\alpha$ parameter change between two consecutive iterations is smaller than $10^-6$. In this example we use the information from the previous section to fix the hyper parameters to values for which the BL solutions tend to the ML ones; we used the hyper parameter values $d=e=0.001$, $a=1$ and $b=c=0.001$ for BL1 and $w_0=1$, $w_1=w_2=0$ for BL2. To asses the quality of the estimators we computed KL-divergences between the true distribution and the estimated ones. Then, for each sample size (N), we performed a paired t-test between the KL-divergences of each possible pair of algorithms and found no statistically significant differences between them. This is not a surprising result since both ML algorithms converge to the same solution and the priors in the Bayesian setting are chosen to tend to obtain a estimator tending to the ML one. In each sub-figure of figure \ref{fig:bias} we present error bars (mean and standard deviation) of the bias in the estimation of each parameter, shape in first row, scale in the second one. Each algorithm is presented in a different column and the x-axis represent the number of samples used for the estimation. These results are, for the ML case, in agreement with the ones presented in \cite{RePEc:vic:vicewp:0908}; the ML shape estimators are upwards biased for the shape and downwards biased for the scale parameter. The bias of the presented Bayesian algorithms is equivalent to that of the existing ML algorithms.   

\begin{figure}[!]
\centering  	
\includegraphics[width=.9\textwidth]{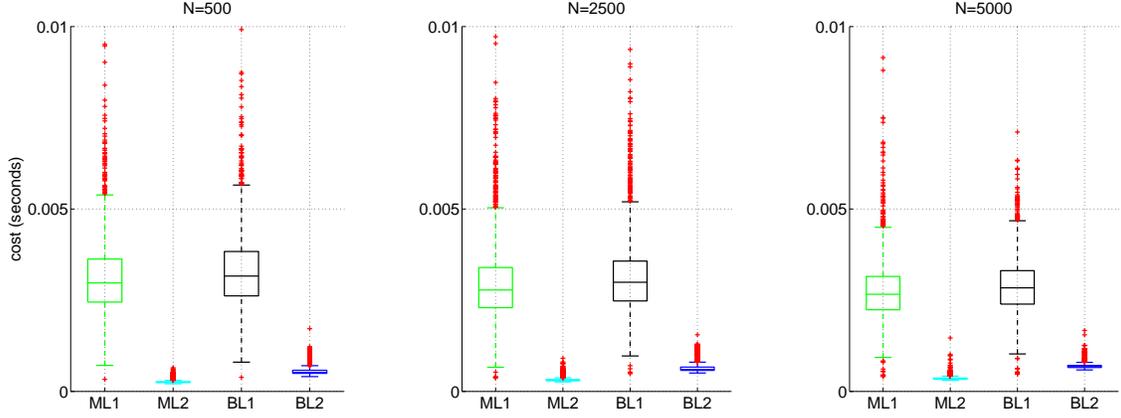}.
\caption{Boxplots of the computational time required by the different algorithms expressed in seconds. Each column represent a different sample size.} 
\label{fig:err}
\end{figure} 
Figure \ref{fig:err} shows boxplots of the computational costs in seconds; the mean, $25^{-th}$ and $75^{-th}$ percentile are presented on each box and the red crosses represent outliers of the estimated distributions. The cost of ML1 and BL1 is equivalent while ML2 and BL2 are much faster. It is to note that in all cases the iterative process depends only on the inverse of the digamma function of a real number and there is no algorithmic cost $N$ dependency in the loop. 


\subsection{Shape conjugate prior hyper parameters values}
\label{res:sh}
It is easy to see flat prior for the shape parameter in the BL2 case can be obtained by choosing $w_1=w_2=0$ in equation (\ref{log_cp}). On the other side, looking at equation (\ref{IG_shape_prior}) it is clear that choosing a flat prior for BL1 is not straightforward. As we showed in section \ref{res:sh}, choosing $b=c \rightarrow 0^+$ and $a \in \mathcal{R}^+$ results in a prior providing a small bias correction which tends to zero for high sample sizes.
To get some more intuition we next provide some numerical examples illustrating such prior for different hyper parameter values and demonstrating the behavior of the posterior hyper parameter updates (equations  \ref{ha},\ref{lha}) and of the Laplace approximation (equation \ref{mu22}) used to obtain the expected shape value $\hat{\alpha}$.

\begin{figure}[!]
\centering  	
\includegraphics[width=.8\textwidth]{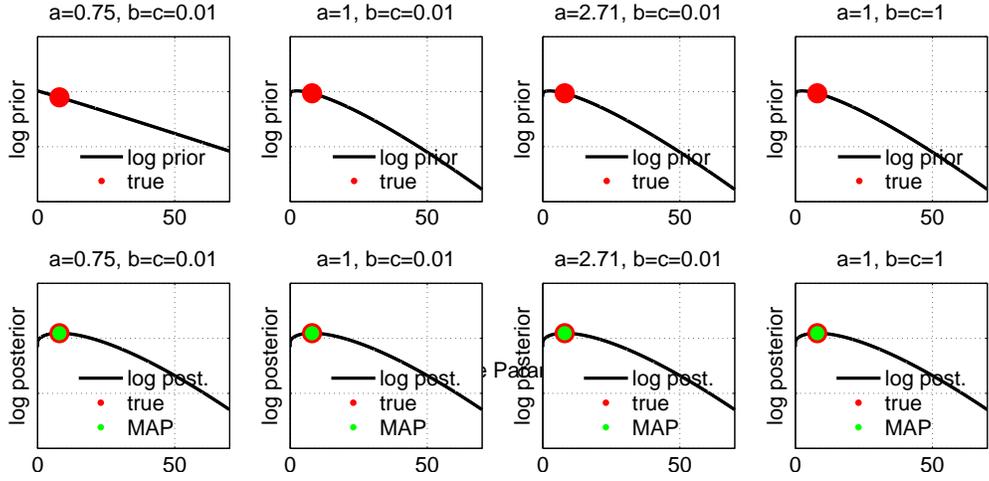}.
\caption{Top row shows the log-prior $\log p(\alpha|a,b,c,\beta)$ and the bottom row the log-posterior $\log p(\alpha|\hat{a},\hat{b},\hat{c},\hat{\beta})$. The red dots mark the shape parameter $\alpha$ from which data was generated. The green dots represent the Bayesian posterior estimation.} 
\label{fig1}
\end{figure}

We generate 1000 samples from an Gamma distribution with parameter values $\alpha=10$ and $\beta=25$. $\hat{\beta}$ is computed through equation (\ref{mmig}) left panel followed by equations (\ref{td},\ref{beta}) where a flat prior was used, namely $d=e=0.01$. For the shape prior hyper parameters we consider different values $a,b,c$, and $\hat{\alpha}$ is computed independently on each case through equations (\ref{ha},\ref{lha},\ref{mu}). In Figure \ref{fig1}, the top row shows the log-prior $\log p(\alpha|a,b,c,\hat{\beta})$ for the parameter values $a,b,c$ indicated in the titles as a function of the shape value $\alpha$ represented in the x-axis. The bottom row shows the corresponding log-posteriors $\log p(\alpha|\hat{a},\hat{b},\hat{c},\hat{\beta})$. In all cases the red dot represents the true shape value, $\alpha=10$, and the green ones in the bottom row represents the posterior estimated value $\hat{\alpha}$.
The expectation we obtained for the posterior of $\alpha$, $\hat{\alpha}$ marked as as green circle, is a very good approximation. Since the parameter value for the scale $\beta$ was also estimated using Bayesian inference and its part of the prior and posterior on $\alpha$, the figure indirectly confirms also the proper behaviour of the scale Bayesian estimation procedure.

These examples are representative of the general behaviour observed for a broad range of hyper parameter values $\{a,b,c\}$. Although this shows the weak dependence on the prior when estimating the posterior, further analyses relating the effect of the hyper parameter value of $a$ with relation to the well known bias correction strategies for the parameters of the Gamma distribution \cite{RePEc:vic:vicewp:0908} would be of interest and are focus of ongoing research. 


\section{Discussion}
\label{dis}
We reviewed two maximum likelihood algorithms for the estimation of the parameters of a Gamma distribution and developed two Bayesian algorithms; the first one uses an unnormalized conjugate prior to the shape parameter and the second one uses a likelihood approximation and prior conjugated with the approximation. In both cases we use the Laplace approximation to compute the required expectations. We performed a theoretical analyses which showed which hyper-parameter values for the conjugate priors will recover the ML algorithms and we also show numerically that with such hyper-parameter choices the new Bayesian algorithms converge towards the ML solutions. We also evaluated numerically the bias in the parameters estimation and show that the Bayesian algorithms have the same bias properties as the ML ones. Including bias corrections in the Bayesian setting for small sample sizes could be achieved as showed in \cite{RePEc:vic:vicewp:0908} for the ML case and is part of ongoing research.    
The Bayesian Learning approaches introduced in section \ref{bayes} has important implications since it can be used inside more complex Bayesian inference tasks, such as fitting mixture models involving Gamma distributions within a (Variational) Bayesian framework to be used for example in the context of medical image segmentation \cite{woolMM}. Extending this beyond modeling one dimensional data, the presented methodology can also be included in multivariate models, e.g. for Bayesian ICA or Non-negative matrix factorizations.


\bibliographystyle{abbrv}
\bibliography{Gamma_InvGm.bib}

\begin{thebibliography}{10}

\bibitem{Lit:Gm_2}
H.~Aksoy.
\newblock Use of gamma distribution in hydrological analysis.
\newblock {\em Journal of Engineering Enviromental Science}, (24):419--228,
  2000.

\bibitem{bishop}
C.~M. Bishop.
\newblock {\em Pattern Recognition and Machine Learning (Information Science
  and Statistics)}.
\newblock Springer, 1 edition, 2007.

\bibitem{cook2008inverse}
J.~D. Cook.
\newblock Inverse gamma distribution.
\newblock {\em Online: http://www. johndcook. com/inverse gamma. pdf,
  Technical. Report}, 2008.

\bibitem{Fink97acompendium}
D.~Fink.
\newblock A compendium of conjugate priors, 1997.

\bibitem{RePEc:vic:vicewp:0908}
D.~Giles and H.~Feng.
\newblock Bias of the maximum likelihood estimators of the two-parameter gamma
  distribution revisited.
\newblock Econometrics Working Papers 0908, Department of Economics, University
  of Victoria, 2009.

\bibitem{Lit:Gm-N-Gm_1}
A.~Khalili, D.~Potter, P.~Yan, L.~Li, J.~Gray, T.~Huang, and S.~Lin.
\newblock Gamma-normal-gamma mixture model for detecting differentially
  methylated loci in three breast cancer cell lines.
\newblock {\em Cancer Inform}, 3:43--54, 2007.

\bibitem{MB}
D.~J.~C. MacKay.
\newblock {\em Information Theory, Inference and Learning Algorithms}.
\newblock Cambrdige University Press, 2003.

\bibitem{Minka_newton}
T.~P. Minka.
\newblock {Beyond Newton's method}.
\newblock Aug. 2000.

\bibitem{minka2002estimating}
T.~P. Minka.
\newblock Estimating a gamma distribution.
\newblock {\em Microsoft Research}, 2002.

\bibitem{woolMM}
M.~Woolrich, T.~Behrens, C.~Beckmann, and S.~Smith.
\newblock Mixture models with adaptive spatial regularization for segmentation
  with an application to fmri data.
\newblock {\em Medical Imaging, IEEE Transactions on}, 24(1):1--11, Jan 2005.

\bibitem{Lit:Gm_1}
S.~Yue.
\newblock A bivariate gamma distribution for use in multivariate flood
  frequency analysis.
\newblock {\em Hydrological Processes}, 15(6):1033--1045, 2001.

\end{thebibliography}

\newpage
\section{Appendices}
\appendix
\section{Method of Moments (MM)}
\label{MM}
 The first two moments of a Gamma distribution are \cite{cook2008inverse}
\begin{align}
\label{moments}
 \mathbb{E}_{\mathcal{G}}[x]=\alpha \beta,   \hspace{2 cm} \mathbb{E}_{\mathcal{G}}[(x-\mathbb{E}_{\mathcal{G}}[x])^2]= \alpha \beta^2.
\end{align}
Using a Gaussian approximation to equation (\ref{moments}) 
\begin{align*}
\mu \approx \alpha \beta,  \hspace{2cm} v \approx \alpha \beta^2, 
\end{align*}
where $\mu$ and $v$ are the mean and variance estimated from the observed data vector $\textbf{x}=\{x_1,\ldots,x_n \}$, and solving the linear system for $\alpha$ and $\beta$ we obtain the MM estimation for the Gamma distribution parameters, namely
\begin{align}
\label{mmig}
\hat{\alpha} \approx  \frac{\mu^2}{v},  \hspace{2 cm} \hat{\beta} \approx \frac{v}{\mu}
\end{align}
Algorithm 3 summarizes the process.
\begin{algorithm}
\label{a4}
\caption{MM for Gamma}
\begin{algorithmic}
\REQUIRE $\textbf{x}=\{x_1,\ldots,x_n \}, x_i > 0$ 
\STATE  $\mu=\frac{1}{n} \sum_{i=1}^{n} x_i$ 
\STATE  $v= \frac{1}{n-1} \sum_{i=1}^{n} (x_i - \mu)^2$
\STATE $\hat{\alpha} =  \frac{\mu^2}{v}$
\STATE $\hat{\beta} = \frac{v}{\mu}$
\RETURN $\hat{\alpha}$,$\hat{\beta}$
\end{algorithmic}
\end{algorithm}

Obviously this approximation is very fast since it simply requires estimation of mean and variance. Nevertheless it will provide more biased estimations as the underlying Gamma distribution deviates more from a Gaussian i.e. as the Gamma distribution higher order moments more deviate from zero.

\section{Maximum Likelihood (ML)}
\label{ML}
In this subsection we describe the two algorithms for ML learning of Gamma distributions parameters originally presented in \cite{minka2002estimating}. 
The log-likelihood of the positive vector of observations $\textbf{x}= \{x_1,\ldots,x_n \}$ under the Gamma distribution (\ref{ig}) can be written as
\begin{align}
\label{igll}
\log \mathcal{G}(\textbf{x}|\alpha,\beta) =  n (\alpha-1) \overline{\log\textbf{x}} - n \log \Gamma(\alpha)  - n \alpha \log \beta   - \frac{n\overline{\textbf{x}}}{\beta}
\end{align}
where for easy of notation the upper bar operand denotes the arithmetic mean operator. Finding ML estimations for the parameters is achieved by maximizing equation (\ref{igll}) with respect to the parameters $\{\alpha,\beta \}$; it is easy to verify that equation (\ref{igll}) has a maximum at 
\begin{align}
\label{thML}
\beta = \frac{\overline{\textbf{x}}} {\alpha},
\end{align}
and that direct maximization of equation (\ref{igll}) with respect to $\alpha$ it is not possible. Substituting equation (\ref{thML})  into equation (\ref{igll}) gives
\begin{align}
\label{igll3}
\log \mathcal{G}(\textbf{x}|\alpha) =  n (\alpha-1) \overline{\log\textbf{x}} - n \log \Gamma(\alpha)  - n \alpha \log \overline{\textbf{x}}  + n \alpha \log \alpha  - n\alpha
\end{align}

Direct maximization of equation (\ref{igll3}) with respect to $\alpha$ is (obviously) also not possible. Using a linear constrain around a given value for $\alpha$, denoted as $\alpha_0$, we have
\begin{align}
\label{subst}
\alpha \log (\alpha) \ge (1+\log \alpha_0)(\alpha-\alpha_0) + \alpha_0 \log(\alpha_0).
\end{align}
Substituting (\ref{subst}) into equation (\ref{igll3}) provides a lower bound on the log-likelihood. Differentiating with respect to $\alpha$, equaling to zero and solving for $\alpha$ we have that 
\begin{align}
\label{alphait}
\hat{\alpha}= \Psi^{-1}(  \log \alpha_0 + \overline{\log \textbf{x}} - \log \bar{x} )
\end{align}
where $\Psi$ represents the \emph{digamma function}. For an initial value of $\alpha$ equation (\ref{alphait}) can be iterated till convergence to obtain the desired value $\hat{\alpha}$ and then $\hat{\beta}$ can be computed using equation (\ref{thML}). In this work we initialize the value of $\alpha$ using the method of moments and Algorithm 4 summarizes the process. 
\begin{algorithm}
\label{a5}
\caption{ML1 (linear constrain)}
\begin{algorithmic}
\REQUIRE $\textbf{x}=\{x_1,\ldots,x_n \}, x_i > 0$ 
\STATE  $\mu= \bar{x} = \frac{1}{n} \sum_{i=1}^{n} x_i$
\STATE  $v= \frac{1}{n-1} \sum_{i=1}^{n} (x_i - \mu)^2$
\STATE $\alpha = \frac{\mu^2}{v}$
\REPEAT
\small
\STATE ${\alpha}$ $\leftarrow$ $\Psi^{-1}(  \log \alpha + \overline{\log \textbf{x}} - \log \bar{x} )$
\normalsize
\UNTIL {convergence}
\STATE $\hat{\alpha} = \alpha$
\STATE $\hat{\beta} = \frac{\bar{x}}{\hat{\alpha}}$ 
\RETURN $\hat{\alpha}$,$\hat{\beta}$
\end{algorithmic}
\end{algorithm}

Algorithm 4 provides an accurate ML solution but it suffers from slow convergence. To accelerate the process, one can approximate Gamma log-likelihood presented in equation (\ref{igll3}) by
\begin{align}
\label{approx}
f(\alpha) = k_0 + k_1 \alpha + k_2 \log \alpha.
\end{align}
Taking first and second derivatives of $f(\alpha)$
\begin{align}
f'(\alpha) = k_1 + \frac{k_2}{\alpha}, \hspace{1cm } f''(\alpha) = - \frac{k_2}{\alpha^2}, 
\end{align}
and matching $f(\alpha)$ and its derivatives to $\log \mathcal{G}(\textbf{x}|\alpha)$ and its first two derivatives with respect to $\alpha$ respectively, we obtain values for $k_0,k_1,k_2$. If $f''(\alpha) < 0$, $f(\alpha)$ has a maximum at $\hat{\alpha}$ iff $f'(\hat{\alpha})=0$. This process gives us the update rule
\begin{align}
\frac{1}{\alpha} \leftarrow \frac{1}{\alpha} + \frac{\overline{\log x}-\log \bar{x} + \log \alpha - \Psi(\alpha)}{\alpha^2(\frac{1}{\alpha} - \Psi'(\alpha))}  
\end{align}

As before we use the MM relationship for the shape parameter (equation \ref{mmig}, left side) to initialize $\alpha$. After iteration till convergence to get $\hat{\alpha}$ we use equation ($\ref{thML}$) to get $\hat{\beta}$. Algorithm 5 summarizes the process:

\begin{algorithm}
\label{M2}
\caption{ML2 (non-linear approximation)}
\begin{algorithmic}
\REQUIRE $\textbf{x}=\{x_1,\ldots,x_n \}, x_i > 0$ 
\STATE  $\mu= \bar{x} = \frac{1}{n} \sum_{i=1}^{n} x_i$
\STATE  $v= \frac{1}{n-1} \sum_{i=1}^{n} (x_i - \mu)^2$
\STATE $\alpha = \frac{\mu^2}{v}$
\REPEAT
\STATE $\frac{1}{\alpha}$ $\leftarrow$ $\frac{1}{\alpha} + \frac{\overline{\log x}-\log \bar{x} + \log \alpha - \Psi(\alpha)}{\alpha^2(\frac{1}{\alpha} - \Psi'(\alpha))} $ 
\UNTIL {convergence}
\STATE $\hat{\alpha} = \alpha$ 
\STATE $\hat{\beta} = \frac{\bar{x}}{\hat{\alpha}}$ 
\RETURN $\hat{\alpha}$,$\hat{\beta}$
\end{algorithmic}
\end{algorithm}

The two ML algorithms presented here deviate from the originally presented in \cite{minka2002estimating} in the initialization. While we use the MM for initialization, in the original work a closed form estimation was used.

\end{document}